\documentclass[aps,pre,amsmath,10pt,twocolumn,superscriptaddress]{revtex4-1}
\usepackage[dvipdfmx]{graphicx}
\usepackage{color}
\usepackage{bm}
\usepackage{braket}
\usepackage{amsmath,amssymb}
\usepackage{mathtools}

\usepackage{physics}
\usepackage[normalem]{ulem}

\begin{document}
\title{Polynomial machine learning potential and its application to global structure search in the ternary Cu-Ag-Au alloy}
\author{Atsuto \surname{Seko}}
\email{seko@cms.mtl.kyoto-u.ac.jp}
\affiliation{Department of Materials Science and Engineering, Kyoto University, Kyoto 606-8501, Japan}
\date{\today}

\begin{abstract}
Machine learning potentials (MLPs) have become indispensable for performing accurate large-scale atomistic simulations and predicting crystal structures.
This study introduces the development of a polynomial MLP specifically for the ternary Cu-Ag-Au system.
The MLP is formulated as a polynomial of polynomial invariants that remain unchanged under any rotation.
The polynomial MLP facilitates not only comprehensive global structure searches within the Cu-Ag-Au alloy system but also reliable predictions of a wide variety of properties across the entire composition range. 
The developed MLP supports highly accurate and efficient atomistic simulations, thereby significantly advancing the understanding of the Cu-Ag-Au system. 
Furthermore, the methodology demonstrated in this study can be easily applied to other ternary alloy systems.
\end{abstract}

\maketitle

\section{Introduction}

Machine learning potentials (MLPs) have increasingly been developed and utilized for performing accurate calculations that are computationally prohibitive with density functional theory (DFT) calculation alone, such as large-scale atomistic simulations and crystal structure predictions 
\cite{
Lorenz2004210,
behler2007generalized,
behler2011atom,
han2017deep,
258c531ae5de4f5699e2eec2de51c84f,
PhysRevB.96.014112,
bartok2010gaussian,
PhysRevB.90.104108,
PhysRevX.8.041048,
PhysRevLett.114.096405,
PhysRevB.95.214302,
PhysRevB.90.024101,
PhysRevB.92.054113,
PhysRevMaterials.1.063801,
Thompson2015316,
wood2018extending,
PhysRevMaterials.1.043603,
doi-10.1137-15M1054183,
doi:10.1063/1.5126336,
khorshidi2016amp,
doi-10.1063-1.4930541,
PhysRevB.92.045131,
QUA:QUA24836,
Freitas2022,
PhysRevB.99.014104,
PhysRevLett.120.156001,
PhysRevB.99.064114,
GUBAEV2019148,
Kharabadze2022,
D3CP02817H}.
MLPs are typically trained using extensive datasets generated from DFT calculations and effectively represent the short-range interatomic interactions by incorporating systematic structural features and machine learning techniques, including artificial neural networks, Gaussian process models, and linear regression models.
As a result, MLPs provide greater accuracy than conventional interatomic potentials and achieve significantly improved computational efficiency compared to DFT calculations.

This study develops a polynomial MLP for the ternary Cu-Ag-Au alloy system. 
The potential energy is modeled using polynomial invariants that are invariant under rotational transformations, considering both the atomic species of central atoms and those of neighboring atoms. 
Explicitly accounting for the atomic species of neighboring atoms is crucial for achieving accurate predictions across a wide range of alloy compositions. 
However, the number of model coefficients increases exponentially with the addition of atomic species, making the application of polynomial MLPs to ternary alloy systems more challenging than to binary systems, for which polynomial MLPs have been previously developed \cite{PhysRevB.102.174104, doi:10.1063/5.0129045, wakai2023efficient}.

This study introduces a polynomial MLP that supports robust global structure searches within the ternary Cu-Ag-Au system and provides accurate predictions of various properties for different structural configurations and alloy compositions. 
To accomplish this, the study extends an iterative procedure previously used for developing polynomial MLPs in elemental systems \cite{arXiv:2403.02570}, enabling comprehensive global structure searches. 
This procedure involves global structure searches and MLP refinements, resulting in a polynomial MLP that offers reliable structure searches and accurate property predictions across the entire range of compositions in the ternary Cu-Ag-Au system.

Note that polynomial MLPs have been developed and applied to predict the structures and excess energies of symmetric tilt grain boundary models for individual elements such as Cu, Ag, and Au \cite{PhysRevMaterials.4.123607}. 
These MLPs have been found to predict grain boundary energies and other properties with greater accuracy than the commonly used embedded atom method potentials \cite{PhysRevB.63.224106,doi:10.1080/01418618708204485,Williams_2006,PhysRevB.69.144113}. 

\section{Methodology}
\label{Cu-Ag-Au:Sec-method}

\subsection{Formulation of polynomial MLP}
\label{Cu-Ag-Au:Sec-method-formulation}

A formulation of the polynomial MLP for multi-component systems has been presented in Ref. \cite{PhysRevB.102.174104}.
This formulation models the potential energy as a function of both the atomic species of central atoms and those of neighboring atoms.
Incorporating the atomic species of neighboring atoms can substantially enhance the predictive power for substitutional structures and facilitate the development of MLPs with high accuracy across a wide range of compositions.
The subsequent section provides a concise overview of the polynomial MLP formulation tailored for ternary systems.

\begin{figure}[tbp]
\includegraphics[clip,width=\linewidth]{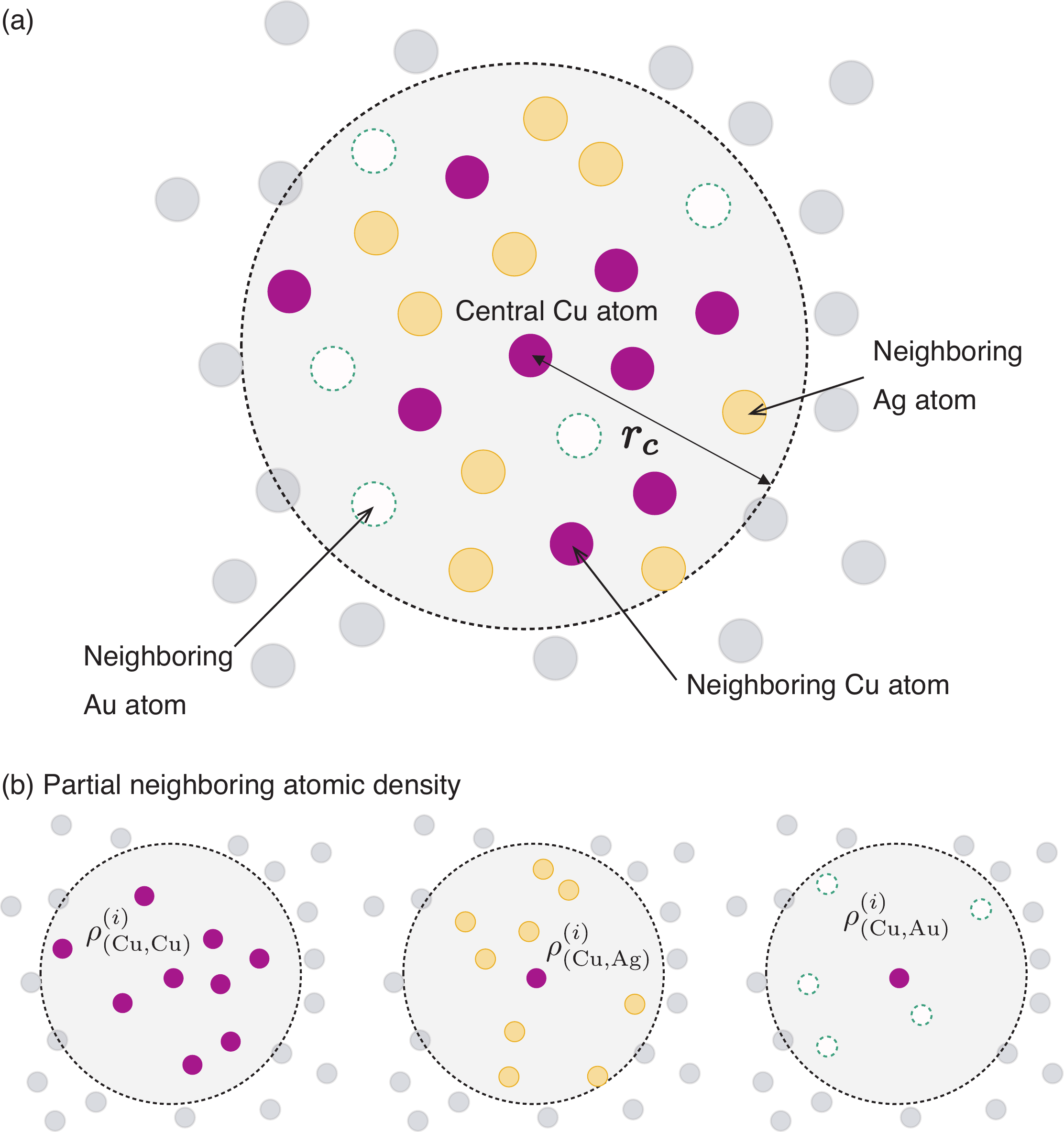}
\caption{
(a) Schematic illustration of the neighboring atomic distribution around atom $i$ of the element Cu within a Cu-Ag-Au ternary structure.
(b) Decomposition of the neighboring atomic density into the partial neighboring atomic densities of the elements Cu, Ag, and Au surrounding atom $i$ of the element Cu.
}
\label{Cu-Ag-Au:atomic-distribution-schematic}
\end{figure}

The polynomial MLP assumes that the short-range part of the potential energy of a structure can be decomposed as the sum of contributions from each atom. 
These contributions are called the atomic energies and are relevant to the neighboring atoms within a given cutoff radius $r_c$. 
The atomic energy of atom $i$ can be approximated using a functional of its partial neighboring atomic densities as, 
\begin{equation}
E^{(i)} = \mathcal{F}_{s_i} \left[ \rho^{(i)}_{(s_i, \rm Cu)} , \rho^{(i)}_{(s_i, \rm Ag)}, \rho^{(i)}_{(s_i, \rm Au)} \right],
\end{equation}
where $\rho^{(i)}_{(s_i,s)}$ denotes the partial neighboring atomic density of element $s$ ($s \in \{{\rm Cu},{\rm Ag},{\rm Au}\}$) around atom $i$ of element $s_i$.
The functional form for the atomic energy should be dependent on the central atom.
Figure \ref{Cu-Ag-Au:atomic-distribution-schematic} provides a schematic illustration of the neighboring atomic distribution around a Cu atom within the cutoff radius $r_c$, and how it decomposes into the partial neighboring atomic densities.

The partial neighboring atomic density of element $s$ around atom $i$ is then expanded in terms of a basis set composed of radial functions $\{f_n\}$ and spherical harmonic functions $\{ Y_{lm} \}$. 
This expansion is expressed as 
\begin{equation}
\rho^{(i)}_{(s_i, s)}(\bm{r}) = \sum_{nlm} a_{nlm,(s_i, s)}^{(i)} f_n(r) Y_{lm} (\bm{r}),
\end{equation}
where $a_{nlm,(s_i,s)}^{(i)}$ denotes the order parameter of component $nlm$ characterizing the partial neighboring atomic density of element $s$ around atom $i$ of element $s_i$.
The current polynomial MLPs adopt a finite set of Gaussian-type radial functions modified by a cosine-based cutoff function to ensure the smooth decay of the radial function \cite{doi:10.1063/5.0129045}.

By introducing a variable for the unordered pair of elements $t$ and defining the order parameter $a_{nlm,t}^{(i)}$ to be zero if $s_i$ is not included in $t$, the atomic energy can be written as a function of the order parameters $\{a_{nlm,t}^{(i)}\}$. 
In the case of the Cu-Ag-Au alloy, the unordered pair $t$ corresponds to any one of the following: $\{\rm Cu,\rm Cu\}$, $\{\rm Cu,\rm Ag\}$, $\{\rm Cu,\rm Au\}$, $\{\rm Ag,\rm Ag\}$, $\{\rm Ag,\rm Au\}$ or $\{\rm Au,\rm Au\}$.

Moreover, an arbitrary rotation leaves the atomic energy invariant \cite{bartok2013representing,PhysRevB.99.214108}.
Therefore, polynomial rotational invariants derived from the order parameters are used to formulate the polynomial MLP.
A $p$th-order polynomial invariant for a radial index $n$ and a set of pairs composed of the angular number and the element unordered pair $\{(l_1,t_1),(l_2,t_2),\cdots,(l_p,t_p)\}$ is given by a linear combination of products of $p$ order parameters, expressed as \cite{PhysRevB.99.214108,PhysRevB.102.174104}
\begin{widetext}
\begin{equation}
\label{tutorial-2022:Eqn-invariant-form}
d_{nl_1l_2\cdots l_p,t_1t_2\cdots t_p,(\sigma)}^{(i)} =
\sum_{m_1,m_2,\cdots, m_p} c^{l_1l_2\cdots l_p,(\sigma)}_{m_1m_2\cdots m_p}
a_{nl_1m_1,t_1}^{(i)} a_{nl_2m_2,t_2}^{(i)} \cdots a_{nl_pm_p,t_p}^{(i)},
\end{equation}
\end{widetext}
where coefficient set $\{c^{l_1l_2\cdots l_p,(\sigma)}_{m_1m_2\cdots m_p}\}$ is independent of the radial index $n$ and the element unordered pair $t$. 
The coefficient set ensures that the linear combinations are invariant for arbitrary rotation.
In terms of fourth- and higher-order polynomial invariants, multiple invariants can be linearly independent for most of the set $\{l_1,l_2,\cdots,l_p\}$, which are distinguished by index $\sigma$ if necessary.
The atomic energy is then modeled as a polynomial function with respect to a given set of polynomial invariants, including a quadratic function \cite{doi:10.1063/5.0129045}.

\subsection{Datasets}
\label{Cu-Ag-Au:Sec-method-dataset}

In this study, a dataset was generated from prototype structures reported for elemental, binary, and ternary systems, as documented in the Inorganic Crystal Structure Database (ICSD) \cite{bergerhoff1987crystal}.
The atomic positions and lattice constants of these structures were optimized using DFT calculations. 
These prototype structures are referred to as ``structure generators''.

Each structure in the dataset was constructed by introducing random lattice expansions, random lattice distortions, and random atomic displacements into a supercell of a structure generator. 
The entire set of generated structures was randomly divided into training and test datasets in a $9:1$ ratio.
Polynomial MLPs were developed using the training dataset, and the prediction errors for energy, force, and stress tensor were evaluated using the test dataset.

Additionally, the dataset includes local minimum structures obtained from random structure searches conducted using the polynomial MLP. 
These local minima were identified during an iterative procedure aimed at creating a polynomial MLP that supports robust random structure searches. 
Details of the iterative procedure for developing the polynomial MLP and performing the random structure search are provided in Sec. \ref{Cu-Ag-Au:Sec-method-go}.

\begin{table*}[tbp]
\begin{ruledtabular}
\caption{
List of prototype structures utilized for generating ternary structures in the training and test datasets. 
The prototype structures are identified by their collection codes and structure types as recorded in the ICSD.
}
\label{Cu-Ag-Au:structure-generators-ternary}
\begin{tabular}{cc|cc|cc|cc}
CollCode & Structure type & CollCode & Structure type & CollCode & Structure type & CollCode & Structure type\\
\hline
456 & Al$_4$Mo$_2$Yb & 30394 & GeSb$_2$Te$_4$ & 67980 & Gd$_2$B$_3$C$_2$ & 189750 & HCaNi$_5$ \\
1133 & Fe$_2$P & 30751 & BaNiSn$_3$ & 68071 & Na$_2$HgO$_2$ & 190705 & CeMg$_2$Si$_2$ \\
2027 & CaGaN & 32619 & Cu$_4$NaAs$_2$ & 68537 & K$_2$PtCl$_6$ & 191257 & BaNiS$_2$ \\
2734 & ThCr$_2$Si$_2$ & 35386 & SmSI & 68798 & Delafossite-NaCrS$_2$ & 192865 & DyNi$_4$Si \\
9564 & Al$_2$Ca$_3$Ge$_2$ & 39452 & Stannite-Cu$_2$FeSnS$_4$ & 71998 & ScAuSi & 201570 & Be$_4$Zr$_2$H$_3$ \\
10011 & Ni$_3$Pb$_2$S$_2$ & 40319 & Ti$_5$Te$_4$ & 75228 & Al$_3$Y$_2$Si$_2$ & 240197 & Fe$_2$Si(HT) \\
10041 & K$_2$CdPb & 40950 & Sc$_3$FeC$_4$ & 76020 & Ti$_3$PO$_2$ & 245196 & Pd$_3$ScH$_2$ \\
10146 & PbClF/Cu$_2$Sb & 41924 & CuSmP$_2$ & 76295 & YAlGe & 246865 & LaGa$_4$ \\
10454 & ZrCuSiAs-CuHfSi$_2$ & 42427 & CoHoGa$_5$ & 78866 & AuEuGe & 391119 & RbCuC$_2$ \\
12150 & La$_2$O$_3$ & 42564 & CuInPt$_2$ & 79005 & FeSi$_4$P$_4$ & 391288 & NdTe$_3$ \\
12157 & KZnAs & 42890 & BaPtSb & 82674 & CeAlSi$_2$ & 407246 & Ce$_2$Si$_2$I$_2$ \\
12163 & K$_3$Cu$_2$P$_2$ & 43034 & CdIn$_2$Se$_4$ & 84825 & Hf$_2$NiP & 409533 & CaLiPb \\
14026 & CrNb$_2$Se$_4$-Cr$_3$S$_4$ & 43843 & Al$_2$Ru$_3$B$_2$ & 87348 & SiU$_3$ & 410967 & Pd$_5$Th$_3$ \\
15128 & Heusler-AlCu$_2$Mn & 44230 & ErIr$_3$B$_2$ & 90460 & FeYbGe & 412038 & KCuC$_2$ \\
15269 & UAsTe & 44293 & Cr$_3$B$_4$ & 93242 & Ti$_2$CuSb$_3$ & 416490 & Gd$_4$Mg$_3$Co$_2$ \\
16203 & CeCr$_2$B$_6$ & 44353 & GaPt$_3$C & 95049 & CePt$_3$B & 417001 & SrFe$_2$As$_2$ \\
16324 & NdBr$_3$ & 44816 & CuHg$_2$Ti & 95072 & U$_4$S$_3$ & 418529 & CeRe$_4$Si$_2$ \\
16475 & Heusler(alloy)-AlLiSi & 44926 & Fe$_2$Tb & 95826 & Sr$_2$AuN & 602022 & FeHTi \\
16501 & BaCuSn$_2$-CeNi$_{(1-x)}$Si$_2$ & 53505 & FeNiN & 99139 & HgNa & 605273 & AgYbS$_2$ \\
16777 & UBC & 53575 & Al$_2$CdS$_4$ & 99161 & Li$_5$Tl$_2$ & 605279 & AgDyTe$_2$ \\
20083 & AlFe$_2$B$_2$ & 54387 & AuBe$_5$ & 100696 & FeTiH$_2$ & 608538 & Cu$_2$HgI$_4$ \\
20320 & LiTiS$_2$ & 55495 & Au$_3$K$_2$ & 102057 & AlRe & 610765 & KSnAs \\
20397 & CeNiC$_2$ & 55570 & CeCoC$_2$ & 102429 & NdNiGa$_2$ & 622688 & Co$_2$In$_4$Lu$_3$ \\
20632 & CeAlCo & 57067 & Ce$_2$CuGe$_6$ & 102444 & Co$_2$InTb & 628479 & Cu$_3$VS$_4$ \\
20876 & InMg$_2$ & 58046 & Al$_5$PrNi$_2$ & 106826 & SrMgIn$_3$ & 629491 & H$_2$FeTi \\
23188 & Mo$_2$BC & 58084 & Al$_5$Ni$_2$Zr & 108776 & Ca$_7$Ge & 631720 & K$_2$MgF$_4$ \\
23255 & AlCe & 58139 & Na$_3$As & 109116 & KCoO$_2$($tP16$) & 632440 & LiFeO$_2$-$\alpha$ \\
23257 & Al$_3$Ti & 60739 & LiMnAs & 150572 & AuCuZn$_2$ & 633208 & KFeS$_2$ \\
23325 & Cu$_4$KS$_3$ & 60829 & LaPtSi & 153861 & Gd$_2$C$_2$Br$_2$ & 635524 & InTe-Tl$_2$Se$_2$ \\
23540 & Pd$_5$TlAs & 61687 & CoYC($tP6$) & 157837 & Ca$_2$CoO$_3$(CoO$_2$)$_{1.61}$ & 637335 & NdSi$_{2-x}$ \\
23550 & Co$_3$GdB$_2$ & 62083 & Cr$_2$Ho$_2$C$_3$ & 160916 & PrCo$_5$H$_x$-frame & 637823 & Bi$_2$Pb$_2$Se$_5$ \\
23586 & Perovskite-CaTiO$_3$ & 62292 & Co$_2$Ge$_4$Tb$_3$ & 167656 & (Ca$_8$)$_x$Ca$_2$ & 638357 & Na$_2$PtH$_4$ \\
23791 & Nb$_2$S$_2$C & 62598 & CoUC$_2$ & 173685 & Ba$_3$Cd$_2$Sb$_4$ & 648539 & InTaS$_2$ \\
25310 & GaGeLi & 63035 & LiPr$_2$Ge$_6$ & 182050 & SrCu$_2$Sn$_2$ & 655086 & Tl$_3$VS$_4$ \\
26284 & MoNiP$_2$ & 66316 & NdRuSi$_2$ & 182349 & Ta$_2$BN$_3$ & 655261 & PdTa$_2$S$_6$ \\
29284 & TiAs & & & & & &
\end{tabular}
\end{ruledtabular}
\end{table*}

To generate structures for the elemental systems of Cu, Ag, and Au, a total of 86 prototype structures were selected, representing single elements with zero oxidation state.
These prototype structures include metallic close-packed structures, covalent structures, layered structures, and high-pressure phase structures. 
A comprehensive list of these structure generators is provided in Ref. \cite{PhysRevB.99.214108}.
For the binary systems Cu-Ag, Cu-Au, and Ag-Au, 150 structure generators were prepared based on the binary prototype structures detailed in Ref. \onlinecite{PhysRevB.102.174104}.
In the case of the ternary Cu-Ag-Au system, a total of 190 prototype structures were selected from those reported as ternary alloy entries in the ICSD. 
A partial list of these ternary prototype structures is presented in Table \ref{Cu-Ag-Au:structure-generators-ternary}, showing only representative prototype structures for which structure types are reported in the ICSD.
Additionally, structures generated by swapping elements within each prototype structure were also considered. 
Consequently, the total number of ternary structure generators amounts to 656.

The dataset currently generated from these structure generators includes a total of 117477 structures. 
Among them, there are 12878 structures for the elemental Cu, 
12874 structures for the elemental Ag, 
12632 structures for the elemental Au, 
11202 structures for the binary Cu-Ag, 
11405 structures for the binary Cu-Au, 
11787 structures for the binary Ag-Au, 
and 44699 structures for the ternary Cu-Ag-Au.

DFT calculations were performed for the structures in the datasets using the plane-wave-basis projector augmented wave method \cite{PAW1} within the Perdew--Burke--Ernzerhof exchange-correlation functional \cite{GGA:PBE96} as implemented in the \textsc{vasp} code \cite{VASP1,VASP2,PAW2}.
The valence electron configurations of Cu, Ag, and Au were $3d^{10}4s^1$, $4d^{10}5s^1$, and $5d^{10}6s^1$, respectively.
The cutoff energy was set to 300 eV.
The allowed spacing between $k$-points was approximately set to 0.09 \AA$^{-1}$.
The total energies converged to less than 10$^{-3}$ meV/supercell.
The atomic positions and lattice constants of the structure generators were optimized until the residual forces were less than 10$^{-2}$ eV/\AA.

\subsection{MLP estimation}
\label{tutorial-2022:Sec-regression}

Polynomial MLPs were developed for universal application across the entire composition range of the Cu-Ag-Au ternary system, utilizing a dataset comprising 117,477 structures.
The coefficients for the polynomial MLPs were estimated from total energy values and force components through weighted linear ridge regression using \textsc{pypolymlp} developed by the author \cite{pypolymlp}. 
Since the forces acting on atoms are described linearly with coefficients identical to those used for the potential energy, the predictor matrix $\bm{X}$ and the observation vector $\bm{y}$ employed in the regression can be represented in submatrix form as 
\begin{equation}
\bm{X} =
\begin{bmatrix}
\bm{X}_{\rm energy} \\
\bm{X}_{\rm force} \\
\end{bmatrix}
,\qquad \bm{y} =
\begin{bmatrix}
\bm{y}_{\rm energy} \\
\bm{y}_{\rm force} \\
\end{bmatrix}.
\end{equation}
The predictor matrix $\bm{X}$ is composed of two submatrices: $\bm{X}_{\rm energy}$ and $\bm{X}_{\rm force}$. 
The former contains polynomial invariants and their products, while the elements of the latter $\bm{X}_{\rm force}$ correspond to the derivatives of the polynomial invariants and their products, as specified in Ref. \onlinecite{PhysRevB.99.214108}.
The observation vector $\bm{y}$ also has two components: $\bm{y}_{\rm energy}$ and $\bm{y}_{\rm force}$. 
These components contain the total energy and the forces acting on atoms for the structures in the training dataset, respectively, which were obtained from DFT calculations.

The dataset used in this study encompasses a substantial number of entries, totaling up to 29,586,198, which includes total energy values and force components across 117,478 distinct structures. 
In some cases, the number of coefficients, i.e., the number of columns in the matrix $\bm{X}$, can exceed 100,000, potentially requiring up to 24 TB of memory to allocate the entire matrix.
To estimate the coefficients of the polynomial MLPs efficiently, this study employs a sequential implementation of linear ridge regression, significantly reducing the memory requirements.
This approach leverages the fact that linear ridge regression involves evaluating $\bm{X}^\top \bm{X}$ and $\bm{X}^\top \bm{y}$ (or their weighted forms) without needing to construct the entire predictor matrix $\bm{X}$.
Specifically, the training dataset is divided into smaller subsets, allowing $\bm{X}^\top \bm{X}$ and $\bm{X}^\top \bm{y}$ to be computed using the relations $\bm{X}^\top \bm{X} = \sum_i \bm{X}_i^\top \bm{X}_i$ and $\bm{X}^\top \bm{y} = \sum_i \bm{X}_i^\top \bm{y}_i$, where $\bm{X}_i$ and $\bm{y}_i$ denote the predictor submatrix and observation vector of the $i$-th subset of the training dataset, respectively.

The magnitude of the penalty in weighted linear ridge regression was meticulously chosen to minimize the root mean square (RMS) error for the test dataset, thereby ensuring optimal estimation of the regression coefficients. 
Furthermore, to enhance the robustness of the MLPs for essential structures, weights were assigned to data entries based on their values \cite{arXiv:2403.02570}. 
Specifically, smaller weights were allocated to less significant data entries characterized by positive energy values and large absolute force components.
This approach contributes to maintaining the predictive accuracy of the polynomial MLP for essential structures, thereby mitigating the risk of reduced performance.

\subsection{Global structure optimization}
\label{Cu-Ag-Au:Sec-method-go}

In this study, a polynomial MLP was developed to facilitate reliable global structure optimizations, utilizing a dataset that includes local minimum structures obtained from random structure searches \cite{Pickard_2011,hendrix2010introduction}, as explained in Sec. \ref{Cu-Ag-Au:Sec-method-dataset}. 
The polynomial MLP was employed to identify globally stable structures on the convex hull of the formation energy, as well as local minimum structures near the convex hull in the Cu-Ag-Au ternary system.
This section provides a brief overview of the procedure used to develop the polynomial MLP and explains its application in finding both global and local minimum structures across a broad range of feasible compositions.

To integrate random structure searches with the polynomial MLP, this study adopts an iterative procedure involving the updating of the polynomial MLPs \cite{arXiv:2403.02570}. This iterative procedure has been adapted for application to multi-component systems.
The procedure is outlined as follows:
(1) A large number of initial structures ($\sim 10^6$) are randomly and uniformly sampled for all possible binary and ternary compositions, represented with up to twelve atoms.
The feasible region is defined by the metric tensor of lattice basis vectors and the fractional coordinates of the atomic positions with respect to the lattice basis vectors.
The feasible region is reduced by applying the main conditions for defining the Niggli reduced cell \cite{ITA2002}.
(2) Local geometry optimizations are systematically performed on the initial structures using the polynomial MLP.
(3) Duplicate local minimum structures are removed using a similarity measure that is relevant to the polynomial MLP \cite{arXiv:2403.02570}.
(4) Single-point DFT calculations are performed for local minimum structures close to the convex hull of the formation energy measured from the energy values of the local minimum structures.
The polynomial MLP is updated by adding their DFT results to the training dataset.

Steps (1)--(4) are iterated until a sufficiently robust random structure search is achieved.
(5) Local geometry optimizations are then performed using DFT calculations, focusing initially on a subset of local minimum structures that are near the convex hull of the formation energy. 
The polynomial MLP may produce small prediction errors that are non-negligible.
Such errors can complicate the assessment of stability between local minimum structures when relying solely on the energy values predicted by the MLP. 
Consequently, DFT calculations are employed in step (5) to accurately determine the final stability of the local minimum structures.

\section{Results and discussion}
\label{Cu-Ag-Au:Sec-results}

\subsection{MLP optimization}

\begin{figure}[tbp]
\includegraphics[clip,width=0.8\linewidth]{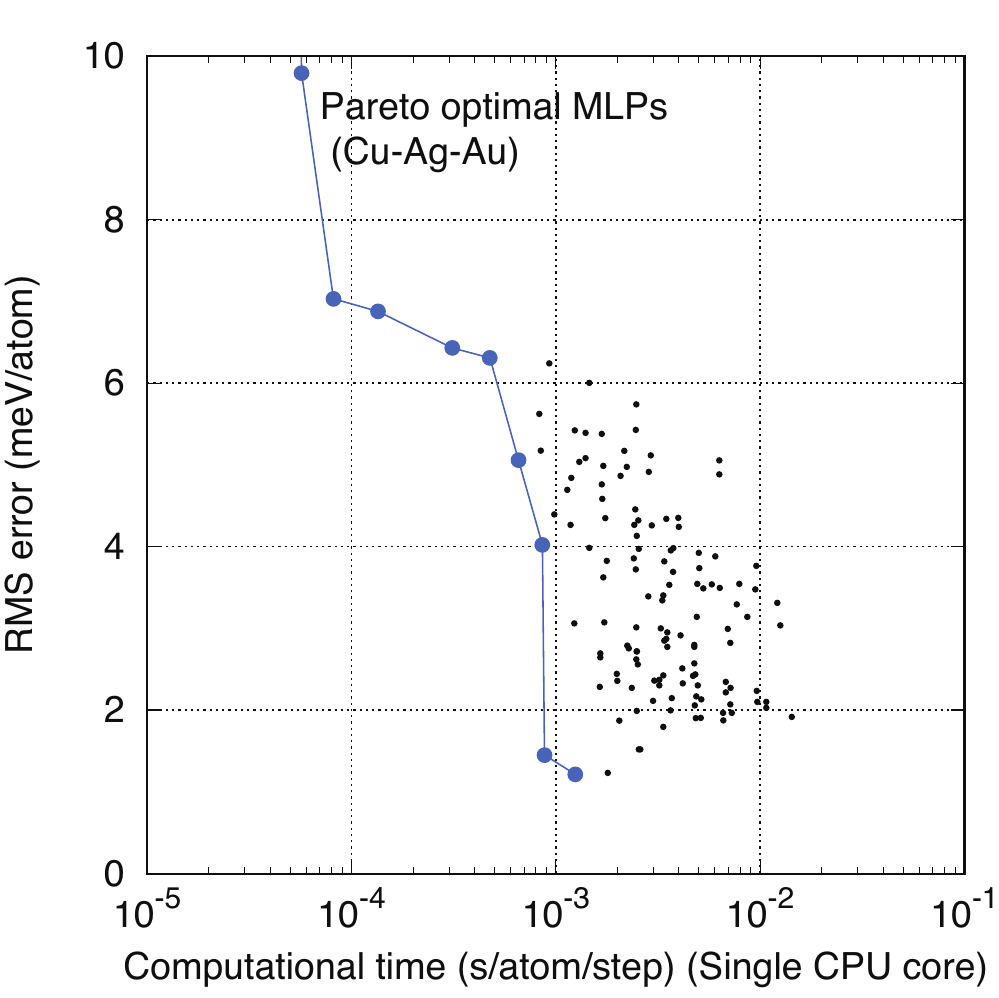}
\caption{
Distribution of polynomial MLPs and Pareto-optimal MLPs for the ternary Cu-Ag-Au system. 
These are obtained from a grid search of input parameters in the polynomial MLP.
The blue closed circles represent the RMS errors of the Pareto-optimal MLPs.
The computational efficiency is assessed by measuring the elapsed time required to compute the energy, forces, and stress tensors of a structure with a large number of atoms. 
This elapsed time is normalized by the number of atoms, as it is proportional to the number of atoms.
The elapsed time for a single point calculation is estimated using a single core of Intel\textregistered\ Xeon\textregistered\ E5-2695 v4 (2.10 GHz) and an implementation of the polynomial MLP to the \textsc{lammps} code \cite{LammpsPolyMLP}.
}
\label{Cu-Ag-Au:Fig-Cu-Ag-Au-pareto}
\end{figure}

\begin{table}[tbp]
\begin{ruledtabular}
\caption{
Model parameters of the selected polynomial MLP in the Cu-Ag-Au system.
}
\label{Cu-Ag-Au:Table-model-parameters}
\begin{tabular}{ll}
RMS error (energy)                         &  1.44 [meV/atom]   \\
RMS error (force)                          &  0.019 [eV/\AA]    \\
Number of coefficients                     &  159165            \\
Cutoff radius                              &  6 [\AA]           \\
Number of radial functions                 &  10                \\
Maximum order of invariants, $p_{\rm max}$ &  2                 \\
Maximum angular number, $l_{\rm max}^{(2)}$ &  4                 \\
Order of polynomial function                &  2                 \\
\end{tabular}
\end{ruledtabular}
\end{table}

The accuracy and computational efficiency of the polynomial MLP model depend on several input parameters \cite{doi:10.1063/5.0129045}. 
These parameters include 
the cutoff radius, 
the type of polynomial function, 
the order of the polynomial function, 
the number of radial functions,
and the truncation of the polynomial invariants, i.e., 
the maximum angular numbers of spherical harmonics $\{l_{\rm max}^{(2)}, l_{\rm max}^{(3)}, \cdots, l_{\rm max}^{(p_{\rm max})}\}$ and the maximum polynomial order of invariants $p_{\rm max}$.
However, the accuracy and computational efficiency are conflicting properties that need to be balanced.
In such cases, Pareto-optimal points can be considered optimal solutions \cite{branke2008multiobjective}.
Therefore, a systematic grid search is conducted to find the Pareto-optimal MLP models.

\begin{figure}[tbp]
\includegraphics[clip,width=0.8\linewidth]{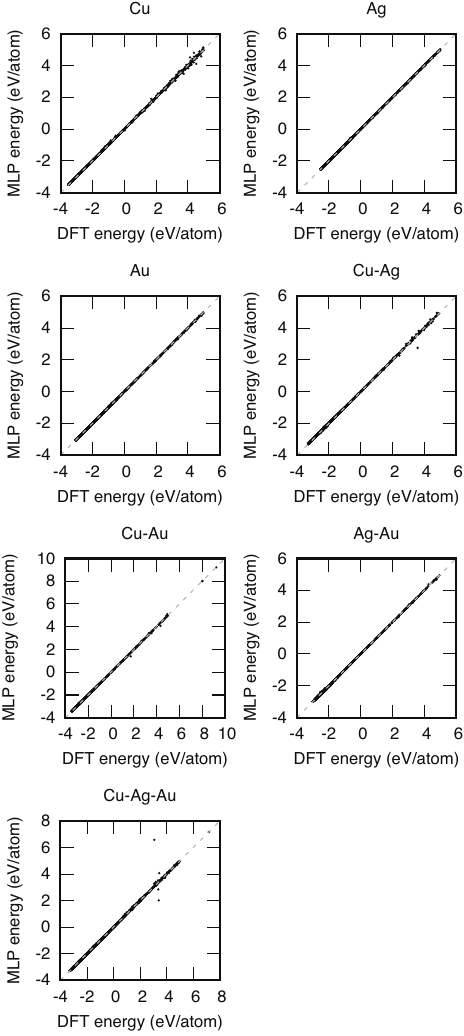}
\caption{
Distribution of energy values for structures included in both the training and test datasets. 
These energy values are predicted using the DFT calculation and the polynomial MLP and are referenced relative to the energy values of isolated atoms. 
The datasets are categorized into seven groups, corresponding to the elemental, binary, and ternary compositions.
}
\label{Cu-Ag-Au:Fig-Cu-Ag-Au-dist}
\end{figure}

Figure \ref{Cu-Ag-Au:Fig-Cu-Ag-Au-pareto} illustrates the distribution of MLPs and the Pareto-optimal MLPs for the ternary Cu-Ag-Au system, obtained from a grid search of input parameters. 
Based on this distribution, a polynomial MLP is selected for global structure searches, and its predictive performance for various properties is evaluated.
The RMS errors and model parameters of the selected MLP are provided in Table \ref{Cu-Ag-Au:Table-model-parameters}.
The RMS errors are assessed by excluding structures with exceptionally high positive energy values, which offers a more practical measure of the accuracy. 
The RMS errors for predicting energy and force are 1.44 meV/atom and 0.019 eV/\AA, respectively.
The selected polynomial MLP is available at \textsc{Polynomial Machine Learning Potential Repository} \cite{MachineLearningPotentialRepository}.

Figure \ref{Cu-Ag-Au:Fig-Cu-Ag-Au-pareto} displays the distribution of energy values predicted by both DFT calculations and the selected polynomial MLP. 
Despite the broad range of prototype structures, compositions within the ternary system, and numerous local minimum structures predicted by the MLP, the energy values predicted by the MLP closely match those obtained from DFT calculations, except for only a few structures showing extremely high positive energy values.

\subsection{Global structure optimization}

\begin{figure*}[tbp]
\includegraphics[clip,width=0.8\linewidth]{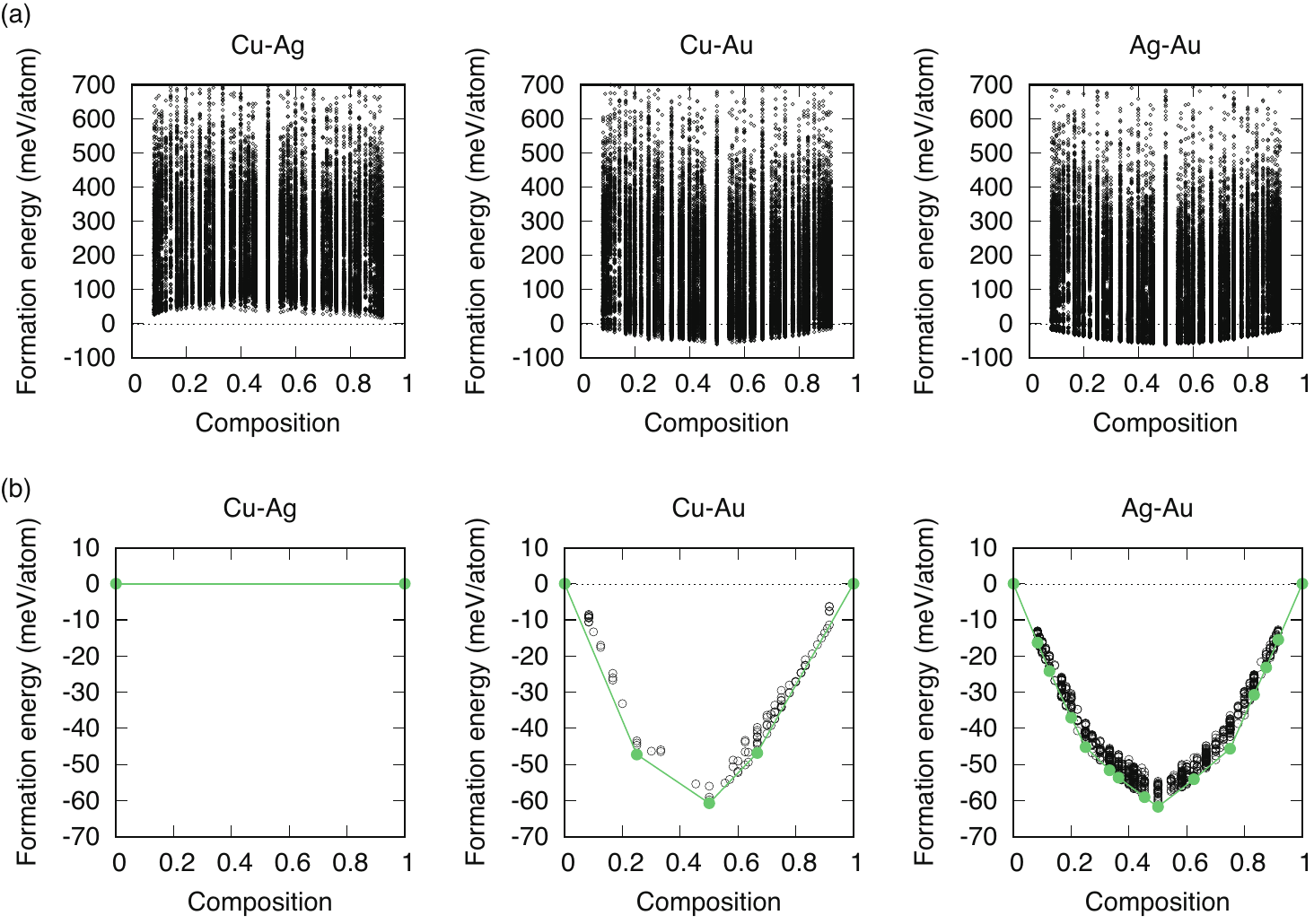}
\caption{
Formation energy values for local minimum structures in the Cu-Ag-Au system at binary compositions.
Panel (a) shows the energy values predicted using the MLP for local minimum structures obtained through random structure searches.
In panel (b),  the formation energy values for local minimum structures close to the convex hull are obtained through the DFT calculation.
The green solid circles represent the stable structures, and the solid lines depict the convex hull of the formation energy for the binary compositions.
}
\label{Cu-Ag-Au:Fig-Cu-Ag-Au-formation-energy}
\end{figure*}

Random structure searches are conducted to identify global minimum structures using the polynomial MLP updated through the iterative procedure described in Section \ref{Cu-Ag-Au:Sec-method-go}.
For 126 binary and ternary compositions, a total of 881,846 initial random structures are sampled, resulting in an extensive number of 21,924,090,668 energy computations using the polynomial MLP. 
Local minimum structures and their formation energy values are subsequently determined through local geometry optimizations starting from these initial structures.

Figure \ref{Cu-Ag-Au:Fig-Cu-Ag-Au-formation-energy} (a) shows the formation energy values of the local minimum structures for binary compositions in Cu-Ag, Cu-Au, and Ag-Au. 
Similarly, the formation energy values are also obtained for ternary compositions in Cu-Ag-Au.
The convex hull of the formation energy is then computed from this dataset, and local minimum structures with formation energy values close to the convex hull are identified as candidates for stable structures.
This study considers structures with energy values relative to the convex hull less than 5 meV/atom (i.e., $\Delta E_{\rm ch}  < 5$ meV/atom).
Consequently, local geometry optimizations are performed for 1431 local structures using DFT calculations. 
Increasing the energy threshold value yields a more reliable convex hull. 
For threshold values of 10, 15, and 20 meV/atom, DFT geometry optimizations are required for 4671, 8565, and 13263 structures, respectively.

\begin{figure*}[tbp]
\includegraphics[clip,width=0.8\linewidth]{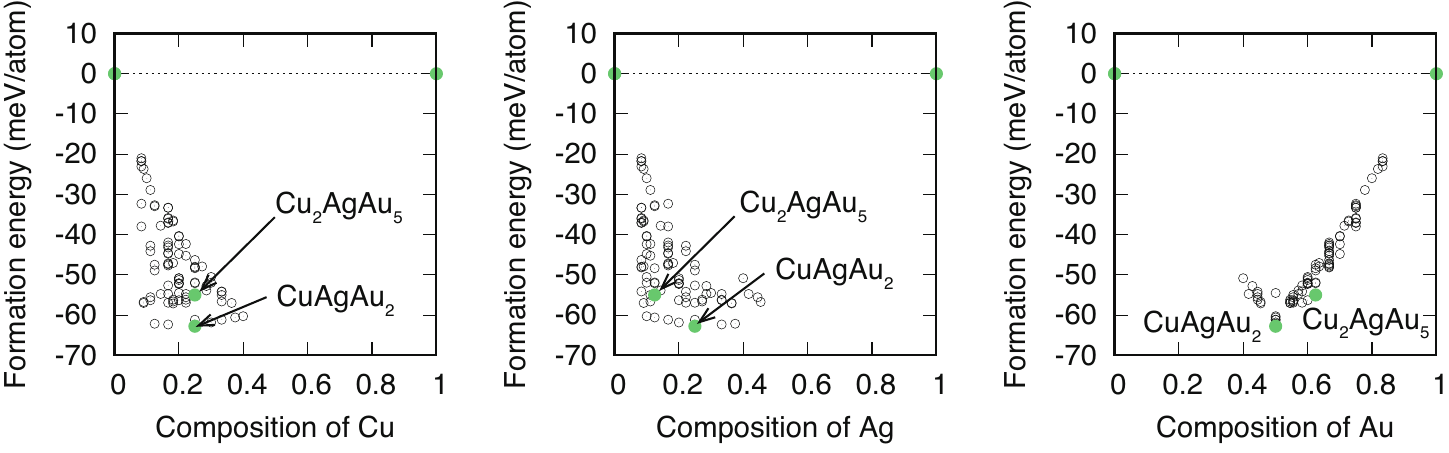}
\caption{
Formation energy values for local minimum structures in the Cu-Ag-Au system at ternary compositions.
The formation energy values for local minimum structures close to the convex hull are accurately obtained using DFT calculations.
The stable structures are represented by the green solid circles.
}
\label{Cu-Ag-Au:Fig-Cu-Ag-Au-formation-energy-ternary}
\end{figure*}

The DFT values of the formation energy are used to identify the convex hull and local minimum structures that are close to the convex hull. 
Figure \ref{Cu-Ag-Au:Fig-Cu-Ag-Au-formation-energy} (b) illustrates the convex hull of the formation energy for binary compositions, along with the formation energy values of local minimum structures obtained through DFT calculations.
Figure \ref{Cu-Ag-Au:Fig-Cu-Ag-Au-formation-energy-ternary} shows the formation energy values of the stable structures on the convex hull and the local minimum structures for ternary compositions. 
These structures are projected onto the compositions of Cu, Ag, and Au.

In the Cu-Ag system, no structures with MLP energy values below the threshold are found. 
As a result, no DFT calculations are performed, and a phase separation state is identified.
Conversely, for binary compositions in the Cu-Au and Ag-Au systems, 3 and 13 structures, respectively, are predicted to lie on the convex hull of the formation energy.
For ternary compositions in the Cu-Ag-Au system, two structures are predicted to be stable.

\begin{table}[tbp]
\begin{ruledtabular}
\caption{
Stable and local minimum structures in binary Cu-Au compositions.
The column labeled $Z$ specifies the number of atoms in the conventional unit cell.
The formation energy and the energy relative to the convex hull are denoted by $\Delta E_f$ and $\Delta E_{\rm ch}$, respectively.
Both energy values are expressed in units of meV/atom.
}
\label{Cu-Ag-Au:Table-structure-list-Cu-Au}
\begin{tabular}{cccccc}
Composition & Structure type & Space group & $Z$ & $\Delta E_f$ & $\Delta E_{\rm ch}$ \\
\hline
\bf Cu$\bf_3$Au  & AuCu$_3$ ($L1_2$)& $ Pm\bar3m $  & 4        & $-47.2$ & 0.0 \\
              & Al$_3$Zr ($D0_{23}$) & $ I4/mmm $ & 16       & $-44.1$ & 3.1 \\
              & Al$_3$Ti ($D0_{22}$) & $ I4/mmm $ & 8        & $-43.4$ & 3.8 \\
Cu$_7$Au$_3$  & Gd$_3$Sn$_7$ & $ Cmmm $ & 20         & $-46.4$ & 3.6 \\
Cu$_2$Au      & UGe$_2$ & $ Cmmm $ & 12           & $-46.5$ & 5.2 \\
              & HfGa$_2$ & $ I4_1/amd $ & 24      & $-46.0$ & 5.8 \\
\bf CuAu      & Rb$_2$Te$_2$ & $ Pbam $ &  8         & $-60.8$ & 0.0 \\
              & IrTa & $ Pmma $ & 12          & $-60.4$ & 0.4 \\
              & IrV & $ Cmmm $ & 8            & $-59.8$ & 0.9 \\
              & AuCu ($L1_0$)& $ P4/mmm $ & 2         & $-59.0$ & 1.7 \\
              & LiSn & $ P2/m $ & 6           & $-56.0$ & 4.7 \\
Cu$_4$Au$_5$  &  $-$  & $ I4/mmm $ &  18         & $-55.1$ & 1.0 \\
Cu$_3$Au$_4$  &  $-$  & $ I4/mmm $ &  14         & $-54.2$ & 0.6 \\
Cu$_2$Au$_3$  &  $-$  & $ I4/mmm $ &  10         & $-52.2$ & 0.1 \\
Cu$_3$Au$_5$  & Pd$_5$Ti$_3$ & $ P4/mmm $ & 8        & $-50.1$ & 0.2 \\
Cu$_4$Au$_7$  &  $-$  & $ I4/mmm $ &  22         & $-49.3$ & 0.0 \\
\bf CuAu$\bf_2$      & CuZr$_2$ & $ I4/mmm $ &  6        & $-46.8$ & 0.0 \\
Cu$_3$Au$_7$  &  $-$  & $ P4/mmm $ &  10         & $-41.5$ & 0.7 \\
Cu$_2$Au$_5$  &  $-$  & $ I4/mmm $ &  14         & $-39.3$ & 0.9 \\
Cu$_3$Au$_8$  &  $-$  & $ I4/mmm $ &  22         & $-37.4$ & 0.9 \\
CuAu$_3$      & $-$  &  $ P4/mmm $ & 12        & $-34.3$ & 0.9 \\
              & Fe$_3$Mn & $ P4/mmm $ & 4         & $-33.9$ & 1.3 \\
Cu$_2$Au$_7$  &  $-$  & $ I4/mmm $ &  18         & $-30.3$ & 0.9 \\
Cu$_2$Au$_8$  &  $-$  & $ P4/nmm $ &  10         & $-27.2$ & 0.9 \\
Cu$_2$Au$_9$  &  $-$  & $ I4/mmm $ &  22         & $-24.7$ & 0.8 \\
CuAu$_5$      & Hf$_5$Pb & $ P4/mmm $ & 6          & $-22.6$ & 0.8 \\
CuAu$_6$      &  $-$  & $ I4/mmm $ &  14         & $-19.4$ & 0.7 \\
CuAu$_7$      &  $-$  & $ P4/mmm $ &  8         & $-16.9$ & 0.7 \\
CuAu$_8$      &  $-$  & $ I4/mmm $ &  18         & $-15.0$ & 0.6 \\
CuAu$_9$      &  $-$  & $ P4/mmm $ &  10         & $-13.6$ & 0.5 \\
CuAu$_{10}$   &  $-$  & $ I4/mmm $ &  22        & $-12.4$ & 0.4 \\
CuAu$_{11}$   &  $-$  & $ P4/mmm $ &  12        & $-11.4$ & 0.3 \\
\end{tabular}
\end{ruledtabular}
\end{table}

Table \ref{Cu-Ag-Au:Table-structure-list-Cu-Au} lists various structures identified through density functional theory DFT calculations for binary compositions of Cu-Au. 
These structures are categorized into three types: those located on the convex hull, local minimum structures with energy values of $\Delta E_{\rm ch}$ less than or equal to 1 meV/atom, and local minimum structures associated with any prototype structures. 
All structures listed have MLP energy values of $\Delta E_{\rm ch}$ less than 5 meV/atom. 
The table also provides the formation energy values and $\Delta E_{\rm ch}$ values obtained from DFT calculations.

The structures located on the convex hull can be found in the compositions of Cu$_3$Au, CuAu, and CuAu$_2$. 
The $L1_2$ structure of Cu$_3$Au is consistent with the experimental structure.
The structure of CuAu$_2$ follows the CuZr$_2$-type structure. 
It is referred to as the ``$\beta2$'' structure, which was predicted to be stable using a combination of DFT calculations and the cluster expansion method \cite{PhysRevB.57.6427}.
The structure of CuAu follows the Rb$_2$Te$_2$-type structure and is not derived from the fcc lattice. 
The predicted structure is not consistent with the experimental structure of the $L1_0$ structure, which shows a slightly larger energy value of 1.7 meV/atom than the ground state.
Additionally, the table contains many face-centered cubic (FCC)-based local minimum structures.

\begin{table}[tbp]
\begin{ruledtabular}
\caption{
Stable and local minimum structures in binary Ag-Au compositions.
The energy values are given in the unit of meV/atom.
}
\label{Cu-Ag-Au:Table-structure-list-Ag-Au}
\begin{tabular}{cccccc}
Composition & Type & Space group & $Z$ & $\Delta E_f$ & $\Delta E_{\rm ch}$ \\
\hline
\bf Ag$_{\bf 11}$Au     &  $-$  & $ Cmmm $ &  24      & $ -16.3 $ & 0.0 \\
Ag$_{10}$Au     &  $-$  & $ C2/m $ &  22      & $ -17.4 $ & 0.4 \\
Ag$_9$Au        &  $-$  & $ C2/m $ &  20      & $ -19.3 $ & 0.2 \\
Ag$_8$Au        &  $-$  & $ C2/m $ &  18      & $ -21.1 $ & 0.5 \\
\bf Ag$\bf_7$Au        &  $-$  & $ Cmmm $ &  16      & $ -24.2 $ & 0.0 \\
                & Ca$_7$Ge & $ Fm\bar3m $ & 32   & $ -22.8 $ & 1.4 \\
Ag$_6$Au        &  $-$  & $ R\bar3 $ &  21     & $ -26.8 $ & 0.4 \\
Ag$_5$Au        &  $-$  & $ Cmmm $ &  24   & $ -31.3 $ & 0.0 \\
                & Al$_5$W  & $ P6_322 $ & 12  & $ -28.6 $ & 2.8 \\
Ag$_9$Au$_2$    &  $-$  & $ C2/m $ &  22   & $ -33.9 $ & 0.1 \\
\bf Ag$\bf_4$Au        & $-$    & $ C2/m $ & 20 & $ -37.1 $ & 0.0 \\
                & MoNi$_4$ ($D1_a$) & $ I4/m $ & 10    & $ -36.8 $ & 0.3 \\
Ag$_7$Au$_2$    &  $-$   & $ C2/m $ &  18   & $ -40.7 $ & 0.0 \\
\bf Ag$\bf_3$Au        & AuCu$_3$ ($L1_2$) & $ Pm\bar3m $ &  4  & $ -45.2 $ & 0.0 \\
                & Al$_3$Zr ($D0_{23}$) & $ I4/mmm $ & 16  & $ -44.9 $ & 0.3 \\
                & Al$_3$Ti ($D0_{22}$) & $ I4/mmm $ & 8  & $ -44.8 $ & 0.5 \\
Ag$_8$Au$_3$    & Al$_8$Mo$_3$ & $ C2/m $ & 22  & $ -46.8 $ & 0.1 \\
Ag$_5$Au$_2$    & Au$_5$Mn$_2$ & $ C2/m $ & 14  & $ -47.6 $ & 0.4 \\
Ag$_7$Au$_3$    & Ga$_3$Pd$_7$ & $ C2/m $ & 20  & $ -48.9 $ & 0.2 \\
\bf Ag$\bf_2$Au        & ReSi$_2$  & $ Immm $ &  6   & $ -51.6 $ & 0.0 \\
\bf Ag$\bf_7$Au$\bf_4$    &  $-$   & $ Immm $ &  22   & $ -53.7 $ & 0.0 \\
Ag$_5$Au$_3$    & Pd$_5$Ti$_3$ & $ P4/mmm $ & 8  & $ -54.3 $ & 0.1 \\
Ag$_3$Au$_2$    &  $-$   & $ Immm $ &  10   & $ -55.6 $ & 0.3 \\
Ag$_7$Au$_5$    &  $-$   & $ Fmmm $ &  48   & $ -56.7 $ & 0.1 \\
Ag$_4$Au$_3$    &  $-$   & $ Immm $ &  14   & $ -56.9 $ & 0.6 \\
Ag$_5$Au$_4$    &  $-$   & $ Immm $ &  18   & $ -58.0 $ & 0.5 \\
\bf Ag$\bf_6$Au$\bf_5$    &  $-$   & $ Immm $ &  22   & $ -59.1 $ & 0.0 \\
\bf AgAu            &  CsI   & $ P4/mmm $ & 2 & $ -61.7 $ & 0.0 \\
                &  PbU   & $ I4_1/amd $ & 8  & $ -58.2 $ & 3.4 \\
Ag$_5$Au$_7$    &  $-$   & $ P2/m $ &  12   & $ -56.3 $ & 0.3 \\
Ag$_2$Au$_3$    &  $-$   & $ C2/m $ &  20   & $ -55.5 $ & 0.1 \\
                & Ga$_3$Ti$_2$ & $ P4/m $ & 10    & $ -55.3 $ & 0.3 \\
\bf Ag$\bf _3$Au$\bf _5$    &  $-$   & $ Cmmm $ & 16    & $ -54.1 $ & 0.0 \\
AgAu$_2$        &  $-$   & $ P2/m $ & 12    & $ -51.0 $ & 0.3 \\
                & UGe$_2$  & $ Cmmm $ & 12    & $ -50.6 $ & 0.6 \\
                & Re$_2$P  & $ Pnma $ & 12    & $ -50.6 $ & 0.6 \\
                & ZrSi$_2$ ($C49$) & $ Cmcm $ & 12    & $ -49.5 $ & 1.8 \\
Ag$_3$Au$_7$    & Ga$_3$Pd$_7$ & $ C2/m $ & 20  & $ -48.7 $ & 0.3 \\
Ag$_2$Au$_5$    & Au$_5$Mn$_2$ & $ C2/m $ & 14  & $ -45.4 $ & 2.7 \\
Ag$_3$Au$_8$    & Al$_8$Mo$_3$ & $ C2/m $ & 22  & $ -44.7 $ & 2.5 \\
\bf AgAu$\bf _3$        & AuCu$_3$ ($L1_2$) & $ Pm\bar3m $ & 4  & $ -45.6 $ & 0.0 \\
                & Al$_3$Ti ($D0_{22}$) & $ I4/mmm $ & 8 & $ -43.4 $ & 2.2 \\
AgAu$_4$        &  $-$   & $ C2/m $ &  20   & $ -36.4 $ & 0.3 \\
                & MoNi$_4$ ($D1_a$) & $ I4/m $ &  10   & $ -34.6 $ & 2.1 \\
\bf AgAu$\bf _5$        &  $-$   & $ Pmma $ &  12   & $ -30.7 $ & 0.0 \\
\bf AgAu$\bf _7$        &  $-$   & $ P4/mmm $ & 8  & $ -23.2 $ & 0.0 \\
AgAu$_8$        & Pt$_8$Ti  & $ I4/mmm $ & 18 & $ -20.4 $ & 0.2 \\
AgAu$_9$        &  $-$   & $ C2/m $ & 20    & $ -18.4 $ & 0.1 \\
AgAu$_{10}$     &  $-$  & $ C2/m $ &  22   & $ -16.4 $ & 0.5 \\
\bf AgAu$\bf _{11}$     &  $-$  & $ Pmmm $ &  12   & $ -15.5 $ & 0.0 \\
\end{tabular}
\end{ruledtabular}
\end{table}

Table \ref{Cu-Ag-Au:Table-structure-list-Ag-Au} presents the stable and local minimum structures identified for binary compositions of Ag-Au.
A substantial number of structures are predicted through random structure searches, with the majority originating from the FCC lattice. 
The FCC solid solution state is only observed above 1173 K and no ordered structures have been reported experimentally.

\begin{table}[tbp]
\begin{ruledtabular}
\caption{
Stable and local minimum structures in ternary Cu-Ag-Au compositions.
The energy values are given in the unit of meV/atom.
}
\label{Cu-Ag-Au:Table-structure-list-Cu-Ag-Au}
\begin{tabular}{cccccc}
Composition & Type & Space group & $Z$ & $\Delta E_f$ & $\Delta E_{\rm ch}$ \\
\hline
Cu$_4$AgAu$_5$ &  $-$  & $ P4/mmm $ & 10        & $ -60.3 $ & 1.3 \\
Cu$_3$AgAu$_4$ &  $-$  & $ P4/mmm $ & 8        & $ -60.6 $ & 1.2 \\
Cu$_4$AgAu$_6$ &  $-$  & $ I4mm $ & 22          & $ -57.0 $ & 2.0 \\
Cu$_2$AgAu$_3$ & MgRu$_2$Sc$_3$ & $ P4/mmm $ & 6     & $ -61.2 $ & 0.9 \\
Cu$_3$AgAu$_5$ &  $-$  & $ I4mm $ & 18          & $ -56.6 $ & 2.0 \\
Cu$_4$AgAu$_7$ &  $-$  & $ P4mm $ & 12          & $ -54.1 $ & 2.8 \\
Cu$_3$Ag$_2$Au$_5$ &  $-$  & $ P4/mmm $ & 10        & $ -61.8 $ & 0.6 \\
Cu$_3$AgAu$_6$ &  $-$  & $ P4mm $ & 10          & $ -51.9 $ & 4.2 \\
Cu$_2$AgAu$_4$ &  $-$  & $ I4mm $ & 14          & $ -53.8 $ & 4.2 \\
Cu$_3$AgAu$_7$ &  $-$  & $ I4mm $ & 22          & $ -48.0 $ & 4.8 \\
{\bf CuAgAu$\bf_2$} & AlRe & $ P4/mmm $ & 4        & $ -62.8 $ & 0.0 \\
{\bf Cu$\bf_2$AgAu$\bf_5$} & Al$_5$Ni$_2$Zr & $ I4/mmm $ & 16   & $ -55.0 $ & 0.0 \\
Cu$_3$AgAu$_8$ &  $-$  & $ C2/m $ & 24          & $ -48.1 $ & 1.2 \\
Cu$_2$Ag$_3$Au$_4$ &  $-$  & $ I4/mmm $ & 18        & $ -56.2 $ & 3.0 \\
Cu$_2$Ag$_2$Au$_5$ &  $-$  & $ I4mm $ & 18          & $ -56.9 $ & 2.3 \\
Cu$_2$AgAu$_6$ &  $-$  & $ I4/mmm $ & 18        & $ -45.2 $ & 4.9 \\
CuAg$_2$Au$_2$ & Er$_2$Mg$_2$Ru & $ I4/mmm $ & 10  & $ -50.8 $ & 5.4 \\
CuAgAu$_3$ &  $-$  & $ Cmcm $ &  20         & $ -56.3 $ & 0.1 \\
Cu$_2$AgAu$_7$ &  $-$  & $ Amm2 $ & 20          & $ -44.8 $ & 1.5 \\
Cu$_2$Ag$_4$Au$_5$ &  $-$  & $ I4mm $ & 22         & $ -57.1 $ & 2.6 \\
Cu$_2$Ag$_3$Au$_6$ &  $-$  & $ I4mm $ & 22          & $ -56.6 $ & 3.0 \\
Cu$_2$Ag$_5$Au$_5$ &  $-$  & $ Pmmm $ & 12          & $ -54.8 $ & 2.5 \\
CuAg$_2$Au$_3$ & MgRu$_2$Sc$_3$ & $ P4/mmm $ & 6   & $ -62.4 $ & 0.0 \\
Cu$_2$Ag$_3$Au$_7$ &  $-$  & $ Pmmm $ & 12          & $ -56.9 $ & 0.4 \\
CuAgAu$_4$ &  $-$  & $ Cmmm $ &  24         & $ -47.7 $ & 4.2 \\
Cu$_2$AgAu$_9$ &  $-$  & $ I4mm $ & 24         & $ -37.1 $ & 1.7 \\
CuAg$_2$Au$_4$ &  $-$  & $ I4mm $ & 14          & $ -54.7 $ & 3.2 \\
CuAg$_3$Au$_4$ &  $-$  & $ P4/mmm $ & 8        & $ -62.1 $ & 0.1 \\
CuAg$_4$Au$_4$ &  $-$  & $ I4mm $ & 18          & $ -55.6 $ & 3.3 \\
CuAg$_3$Au$_5$ &  $-$  & $ I4mm $ & 18          & $ -56.3 $ & 2.5 \\
CuAg$_5$Au$_5$ &  $-$  & $ I4mm $ & 22          & $ -56.8 $ & 2.7 \\
CuAg$_4$Au$_6$ &  $-$  & $ I4mm $ & 22          & $ -57.0 $ & 2.3 \\
CuAg$_2$Au$_9$ &  $-$  & $ P4/mmm $ & 12        & $ -38.0 $ & 4.2 \\
CuAgAu$_{10}$ &  $-$  & $ I\bar4m2 $ & 24     & $ -23.0 $ & 4.1 \\
\end{tabular}
\end{ruledtabular}
\end{table}

Table \ref{Cu-Ag-Au:Table-structure-list-Cu-Ag-Au} provides a list of stable and local minimum structures in ternary Cu-Ag-Au compositions. 
The table shows that many FCC-based local minimum structures are present, as well as the cases of the binary Cu-Au and Ag-Au systems. 
However, no experimental structures have been reported yet. 
Figure \ref{Cu-Ag-Au:Fig-Cu-Ag-Au-formation-energy-ternary} and Table \ref{Cu-Ag-Au:Table-structure-list-Cu-Ag-Au} reveal that two stable structures of CuAgAu$_2$ and Cu$_2$AgAu$_5$ exist, which correspond to (100) superlattices and are identified as the AlRe- and Al$_5$Ni$_2$Zr-types, respectively.
These structures are derived by substituting elements with the Fe$_3$Mn-type structure of CuAu$_3$ denoted by ``$Z3$'' structure in Ref. \onlinecite{PhysRevB.57.6427}, showing $\Delta E_{\rm ch} = 1.3$ meV/atom in Table \ref{Cu-Ag-Au:Table-structure-list-Cu-Au}.

\subsection{Predictive power}

\begin{table}[tbp]
\begin{ruledtabular}
\caption{
Lattice constants and elastic constants for the elemental and binary compounds in the Cu-Ag-Au system.
}
\label{Cu-Ag-Au:Table-lattice-constants}
\begin{tabular}{llcc}
Compound & & MLP & DFT  \\
\hline
Cu (FCC) & $a$ (\AA)      & 3.632 & 3.628 \\
         & $C_{11}$ (GPa) & 180.8 & 177.8 \\
         & $C_{12}$ (GPa) & 122.0 & 125.7 \\
         & $C_{44}$ (GPa) & 77.8  & 80.4  \\
\hline
Ag (FCC) & $a$ (\AA)      & 4.148 & 4.141 \\
         & $C_{11}$ (GPa) & 113.1 & 118.9 \\
         & $C_{12}$ (GPa) & 86.4  & 83.3 \\
         & $C_{44}$ (GPa) & 39.4  & 44.6 \\
\hline
Au (FCC) & $a$ (\AA)      & 4.157 & 4.156 \\
         & $C_{11}$ (GPa) & 153.1 & 157.5 \\
         & $C_{12}$ (GPa) & 138.3 & 133.1 \\
         & $C_{44}$ (GPa) & 25.7  & 27.7  \\
\hline
Cu$_3$Au ($L1_2$) & $a$ (\AA)      & 3.784 & 3.778 \\
                  & $C_{11}$ (GPa) & 179.5 & 181.7 \\
                  & $C_{12}$ (GPa) & 122.0 & 124.6 \\
                  & $C_{44}$ (GPa) & 63.3  & 66.3  \\
\hline
CuAu ($L1_0$)     & $a$ (\AA)      & 2.885 & 2.873 \\
                  & $c$ (\AA)      & 3.607 & 3.630 \\
                  & $C_{11}$ (GPa) & 173.0 & 157.0 \\
                  & $C_{12}$ (GPa) & 116.6 & 120.2 \\
                  & $C_{13}$ (GPa) & 134.5 & 131.0 \\
                  & $C_{33}$ (GPa) & 149.2 & 150.6 \\
                  & $C_{44}$ (GPa) & 59.3  & 61.5  \\
                  & $C_{66}$ (GPa) & 36.0  & 37.6  \\
\hline
CuAu$_2$          & $a$ (\AA)      & 2.887 & 2.880 \\
                  & $c$ (\AA)      & 11.527 & 11.559 \\
                  & $C_{11}$ (GPa) & 178.1 & 162.1 \\
                  & $C_{12}$ (GPa) & 118.5 & 105.7 \\
                  & $C_{13}$ (GPa) & 129.5 & 118.0 \\
                  & $C_{33}$ (GPa) & 150.1 & 136.1 \\
                  & $C_{44}$ (GPa) & 42.8  & 41.6  \\
                  & $C_{66}$ (GPa) & 24.4  & 25.8  \\
\hline
Ag$_3$Au ($L1_2$) & $a$ (\AA)      & 4.147 & 4.141 \\
                  & $C_{11}$ (GPa) & 122.8 & 130.7 \\
                  & $C_{12}$ (GPa) & 93.2  & 93.6  \\
                  & $C_{44}$ (GPa) & 40.3  & 43.2  \\
\hline
AgAu (CsI-type)   & $a$ (\AA)      & 2.918 & 2.910 \\
                  & $c$ (\AA)      & 4.191 & 4.200 \\
                  & $C_{11}$ (GPa) & 156.1 & 162.4 \\
                  & $C_{12}$ (GPa) & 79.4  & 86.0  \\
                  & $C_{13}$ (GPa) & 103.0 & 104.7 \\
                  & $C_{33}$ (GPa) & 139.3 & 142.8 \\
                  & $C_{44}$ (GPa) & 37.6  & 39.3  \\
                  & $C_{66}$ (GPa) & 14.0  & 17.0  \\
\hline
AgAu$_3$ ($L1_2$) & $a$ (\AA)      & 4.151 & 4.148 \\
                  & $C_{11}$ (GPa) & 141.6 & 148.5 \\
                  & $C_{12}$ (GPa) & 115.9 & 119.0 \\
                  & $C_{44}$ (GPa) & 32.8  & 33.1  \\
\end{tabular}
\end{ruledtabular}
\end{table}

\begin{table}[tbp]
\begin{ruledtabular}
\caption{
Lattice constants and elastic constants for the ternary compounds in the Cu-Ag-Au system. 
}
\label{Cu-Ag-Au:Table-lattice-constants2}
\begin{tabular}{llcc}
Compound & & MLP & DFT  \\
\hline
CuAgAu$_2$ & $a$ (\AA)      & 2.908 & 2.906 \\
           & $c$ (\AA)      & 7.769 & 7.759 \\
           & $C_{11}$ (GPa) & 161.9 & 155.6 \\
           & $C_{12}$ (GPa) & 98.1  & 103.5 \\
           & $C_{13}$ (GPa) & 117.2 & 113.5 \\
           & $C_{33}$ (GPa) & 142.2 & 134.3 \\
           & $C_{44}$ (GPa) & 46.9  & 48.9  \\
           & $C_{66}$ (GPa) & 24.3  & 28.8  \\
\hline
Cu$_2$AgAu$_5$ & $a$ (\AA)      & 4.121 & 4.120 \\
               & $c$ (\AA)      & 15.512& 15.474\\
               & $C_{11}$ (GPa) & 159.5 & 167.9 \\
               & $C_{12}$ (GPa) & 114.6 & 113.0 \\
               & $C_{13}$ (GPa) & 124.2 & 127.3 \\
               & $C_{33}$ (GPa) & 145.4 & 145.6 \\
               & $C_{44}$ (GPa) & 42.6  & 43.4  \\
               & $C_{66}$ (GPa) & 29.3  & 25.3  \\
\hline
CuAgAu$_3$     & $a$ (\AA)      & 4.159 & 4.131 \\
               & $b$ (\AA)      & 19.491& 19.612\\
               & $c$ (\AA)      & 4.132 & 4.121 \\
               & $C_{11}$ (GPa) & 154.3 & 153.9 \\
               & $C_{12}$ (GPa) & 118.9 & 116.2 \\
               & $C_{13}$ (GPa) & 108.5 & 102.4 \\
               & $C_{22}$ (GPa) & 141.8 & 143.5 \\
               & $C_{23}$ (GPa) & 119.8 & 116.0 \\
               & $C_{33}$ (GPa) & 152.5 & 152.2 \\
               & $C_{44}$ (GPa) & 43.2  & 43.0  \\
               & $C_{55}$ (GPa) & 27.2  & 28.6  \\
               & $C_{66}$ (GPa) & 41.3  & 43.2  \\
\hline
CuAg$_2$Au$_3$ & $a$ (\AA)      & 2.910 & 2.904 \\
               & $c$ (\AA)      & 11.967& 11.987\\
               & $C_{11}$ (GPa) & 159.5 & 149.7 \\
               & $C_{12}$ (GPa) & 91.5  & 86.6  \\
               & $C_{13}$ (GPa) & 111.2 & 102.2 \\
               & $C_{33}$ (GPa) & 142.8 & 133.4 \\
               & $C_{44}$ (GPa) & 43.4  & 44.2  \\
               & $C_{66}$ (GPa) & 20.7  & 23.7  \\
\hline
CuAg$_3$Au$_4$ & $a$ (\AA)      & 2.912 & 2.907 \\
               & $c$ (\AA)      & 16.155& 16.171\\
               & $C_{11}$ (GPa) & 158.6 & 149.7 \\
               & $C_{12}$ (GPa) & 88.4  & 84.8  \\
               & $C_{13}$ (GPa) & 109.0 & 105.2 \\
               & $C_{33}$ (GPa) & 141.7 & 140.4 \\
               & $C_{44}$ (GPa) & 41.9  & 44.0  \\
               & $C_{66}$ (GPa) & 18.9  & 22.3  \\
\end{tabular}
\end{ruledtabular}
\end{table}

\subsubsection{Equation of states, lattice constants, and elastic constants}

Table \ref{Cu-Ag-Au:Table-lattice-constants} summarizes the lattice constants of various elemental and binary stable compounds within the Cu-Ag-Au system.
This includes FCC structures of Cu, Ag, and Au, as well as the $L1_2$-type Cu$_3$Au, $L1_0$-type CuAu, CuAu$_2$, $L1_2$-type Ag$_3$Au, CsI-type AgAu, and $L1_2$-type AgAu$_3$.
Table \ref{Cu-Ag-Au:Table-lattice-constants2} presents the lattice constants for ternary stable and metastable compounds, specifically CuAgAu$_2$, Cu$_2$AgAu$_5$, CuAgAu$_3$, CuAg$_2$Au$_3$, and CuAg$_3$Au$_4$. 
The lattice constants for these compounds are derived using both polynomial MLP and DFT calculations. 
In most cases, the relative error in lattice constants obtained from MLP is less than 0.3\%.
Furthermore, Tables \ref{Cu-Ag-Au:Table-lattice-constants} and \ref{Cu-Ag-Au:Table-lattice-constants2} also present the elastic constants for the stable and metastable compounds in the Cu-Ag-Au system. 
The relative error in the elastic constants calculated using MLP is within 10\%.
These findings demonstrate that the lattice constants and elastic constants obtained using MLP closely match those determined through DFT calculations.

\begin{figure*}[tbp]
\includegraphics[clip,width=0.9\linewidth]{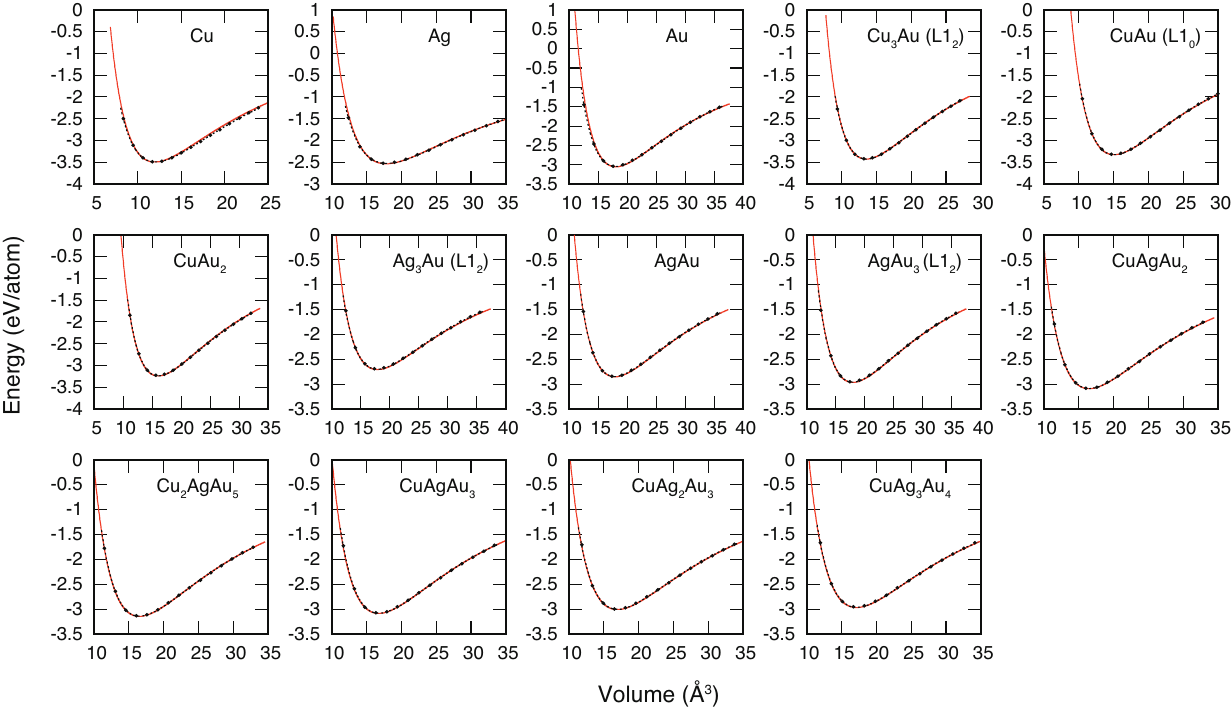}
\caption{
Energy-volume curves for the stable and metastable compounds, which are computed using the polynomial MLP and the DFT calculation.
The red solid line represents the energy-volume curve obtained using the polynomial MLP.
The black closed circles denote the energy values derived from the DFT calculations. 
The black dashed line shows the fitted energy-volume curve corresponding to the DFT calculations.
}
\label{Cu-Ag-Au:Fig-Cu-Ag-Au-eos}
\end{figure*}

Figure \ref{Cu-Ag-Au:Fig-Cu-Ag-Au-eos} displays the energy-volume curves for the stable and metastable compounds, computed using both the polynomial MLP and the DFT calculation.
The equations of state, bulk modulus, and equilibrium volume are determined by fitting a set of volume and energy values to the Vinet equation \cite{https://doi.org/10.1029/JB092iB09p09319}.
The energy-volume curves computed using polynomial MLP are almost identical to those obtained from DFT calculations.
However, a slight deviation is observed between the energy-volume curves derived from MLP and DFT for the elemental Cu, Ag, and Au at small volumes.

\subsubsection{Phonon properties}

\begin{figure}[tbp]
\includegraphics[clip,width=\linewidth]{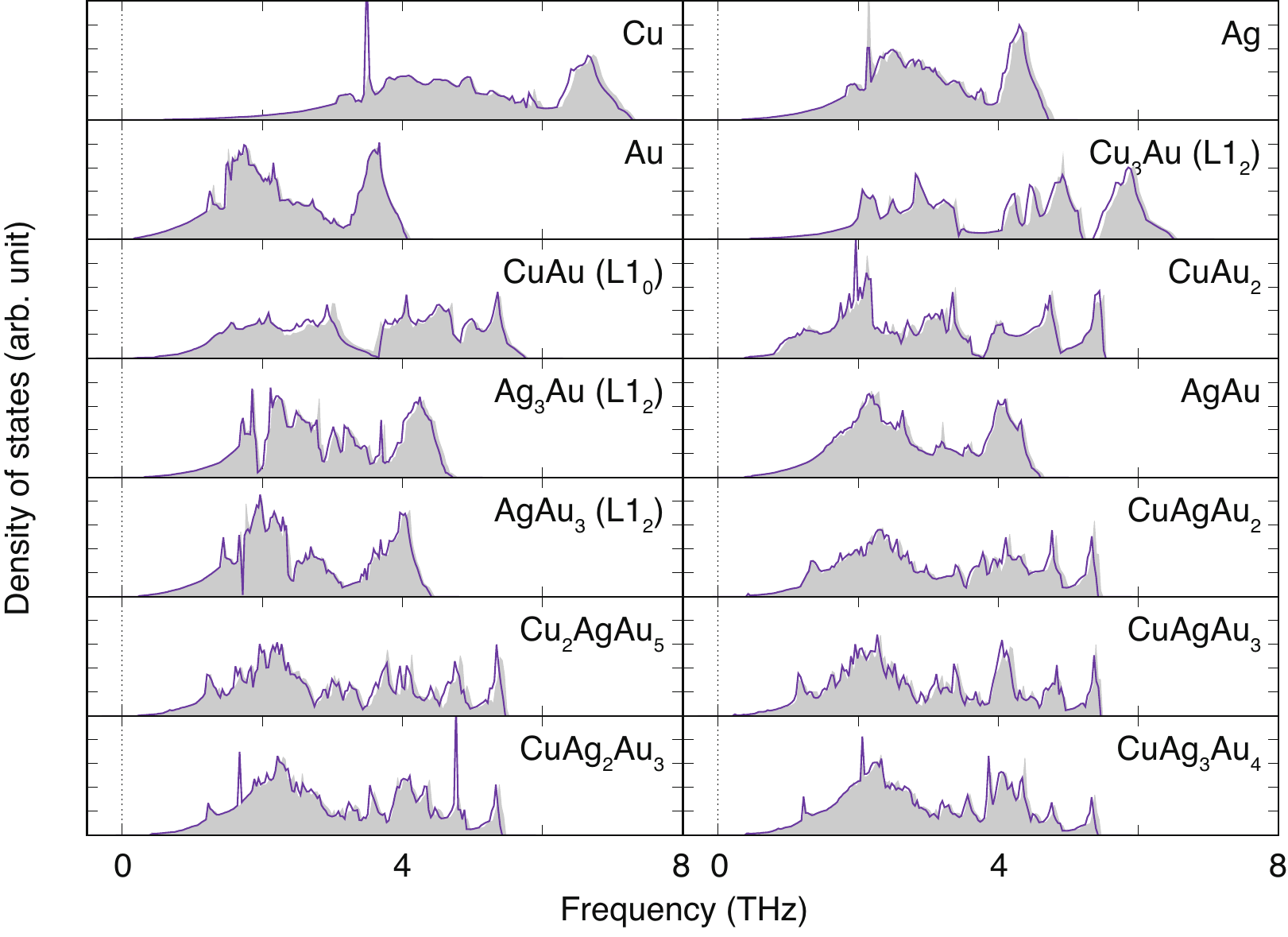}
\caption{
Phonon DOS for the stable and metastable compounds in the Cu-Ag-Au system.
The purple solid line shows the phonon DOS obtained using the polynomial MLP.
The shaded region represents the phonon DOS obtained using the DFT calculation.
}
\label{Cu-Ag-Au:Fig-Cu-Ag-Au-phonon-dos}
\end{figure}

The phonon frequencies of various compounds are calculated using supercell force constants, which are obtained from the polynomial MLP and DFT calculations. 
For each compound, a set of atomic displacements and forces on atoms in supercells is used to estimate supercell force constants. 
To obtain the forces on atoms, a finite displacement of 0.01 \AA\ is applied to a single atom in the equilibrium supercell structure. 
The supercell size is chosen to be as isotropic as possible and sufficiently large to reduce size effects.
Phonon frequency calculations and supercell generation are performed using the \textsc{phonopy} code \cite{togo2015first}. 
Figure \ref{Cu-Ag-Au:Fig-Cu-Ag-Au-phonon-dos} presents the phonon density of states (DOS) for the stable and metastable compounds in the Cu-Ag-Au system. 
The results indicate that the phonon DOS calculated using the polynomial MLP is consistent with that obtained from the DFT calculations for all the compounds.

\subsubsection{Generalized stacking fault energy}

This study employs the following procedure to develop models for calculating generalized stacking fault energy profiles \cite{dumitraschkewitz2017impact}. 
Initially, a supercell consisting of 48 atoms is constructed by expanding the equilibrium structure along the $[111]$ direction of the FCC lattice.
This supercell is then tilted using a displacement vector defined as a linear combination of two vectors perpendicular to the $[111]$ direction, namely $[\bar110]/2$ and $[11\bar2]/2$.
The displacement vector is represented by
\begin{equation}
\bm{b} = \frac{1}{2} u [\bar110] + \frac{1}{2} v [11\bar2],
\end{equation}
where $u$ and $v$ denote the fractional coordinates for the corresponding vectors.

\begin{figure}[tbp]
\includegraphics[clip,width=\linewidth]{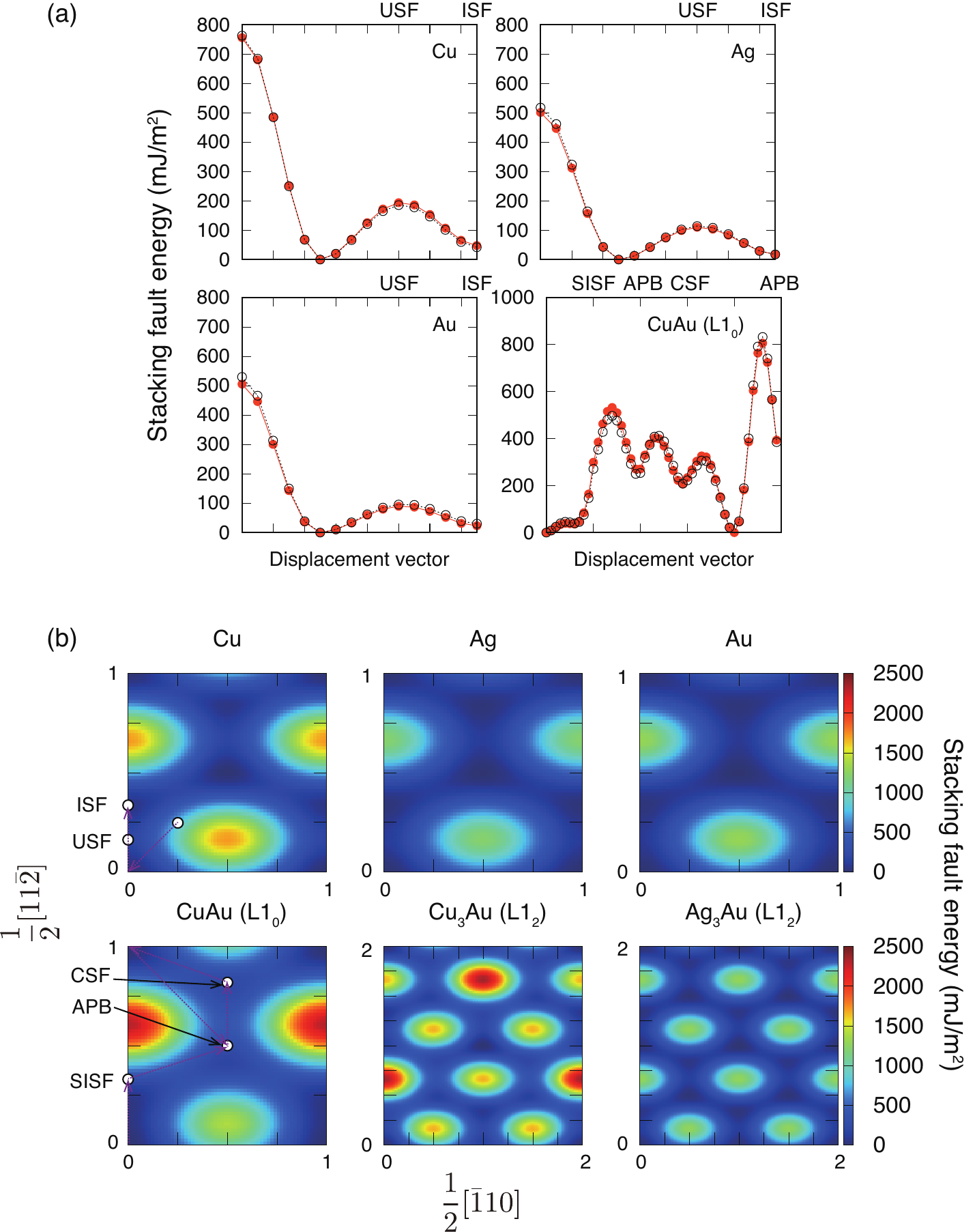}
\caption{
(a) Stacking fault energy profiles in the fcc-Cu, fcc-Ag, fcc-Au, and $L1_0$-type CuAu.
The red closed circles and black open circles represent the stacking fault energy values calculated using the polynomial MLP and DFT calculation, respectively.
(b) Displacement vector dependence of the generalized stacking fault energy for the fcc-Cu, fcc-Ag, fcc-Au, $L1_0$-type CuAu, $L1_2$-type Cu$_3$Au, and $L1_2$-type Ag$_3$Au.
}
\label{Cu-Ag-Au:Fig-Cu-Ag-Au-gsfe}
\end{figure}

Calculation models for generalized stacking faults are systematically constructed, and the excess energy values for these models are obtained from single-point calculations using the polynomial MLP and DFT calculations.
Figure \ref{Cu-Ag-Au:Fig-Cu-Ag-Au-gsfe} (a) shows the stacking fault energy profiles along a path through the intrinsic stacking fault (ISF) and unstable stacking fault (USF) for the elemental Cu, Ag, and Au. 
Figure \ref{Cu-Ag-Au:Fig-Cu-Ag-Au-gsfe} (b) presents the generalized stacking fault energy surface, with the path of the displacement vectors illustrated.
The displacement vectors for the ISF and USF are expressed as 
\begin{eqnarray}
\bm{b}_{\rm ISF} &=& \frac{1}{6} [11\bar2], \nonumber \\
\bm{b}_{\rm USF} &=& \frac{1}{12} [11\bar2].
\end{eqnarray}
The stacking fault energy profiles calculated using the polynomial MLP are almost identical to those calculated using the DFT.
However, the stacking fault energy values are slightly larger than those found in the literature, where DFT calculations include atomic position relaxations along the $[111]$ direction in the tilted structures \cite{Hunter_2013}.

Figure \ref{Cu-Ag-Au:Fig-Cu-Ag-Au-gsfe} (a) also shows the profile of the stacking fault energy along a path through the superlattice intrinsic stacking fault (SISF), the antiphase boundary (APB), and the complex stacking fault (CSF) in the $L1_0$-type CuAu.
Figure \ref{Cu-Ag-Au:Fig-Cu-Ag-Au-gsfe} (b) shows the generalized stacking fault energy surface for the $L1_0$-type CuAu, the $L1_2$-type Cu$_3$Au, and the $L1_2$-type Ag$_3$Au.
The displacement vectors for the SISF, APB, and CSF are represented as
\begin{eqnarray}
\bm{b}_{\rm SISF} &=& \frac{1}{6} [11\bar2], \nonumber \\
\bm{b}_{\rm APB}  &=& \frac{1}{4} [\bar110] + \frac{1}{4} [11\bar2], \\
\bm{b}_{\rm CSF}  &=& \frac{1}{4} [\bar110] + \frac{5}{12} [11\bar2]. \nonumber
\end{eqnarray}
The stacking fault energy profiles calculated using the polynomial MLP and the DFT are consistent. 
These results indicate that the current MLP has high predictive power for the stacking faults and related properties.

\subsubsection{Vacancy formation energy}

\begin{figure}[tbp]
\includegraphics[clip,width=\linewidth]{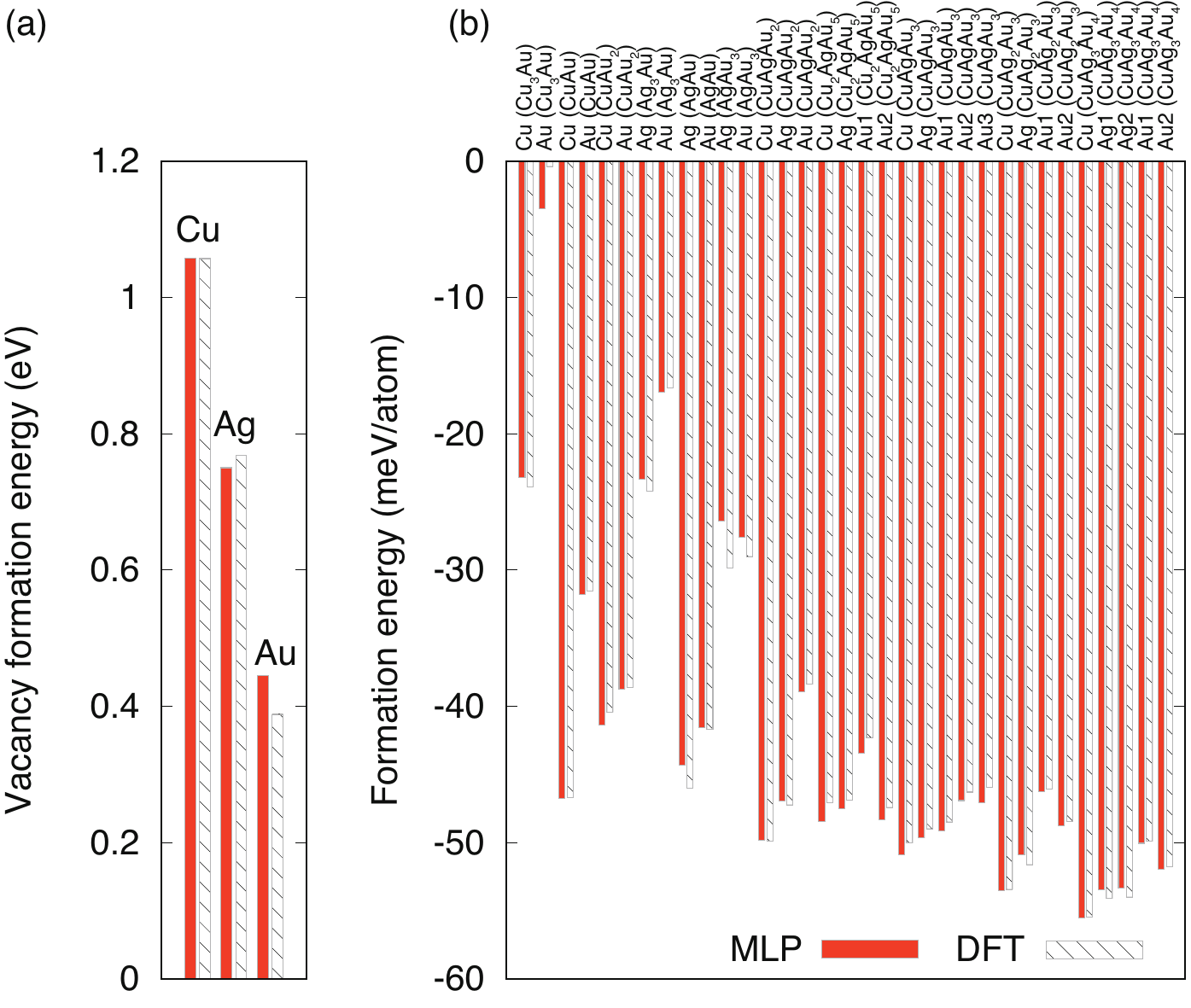}
\caption{
(a) Vacancy formation energy in the elemental Cu, Ag, and Au.
(b) Formation energy for the stable and metastable compounds containing single vacancies. 
The formation energy is measured from the energy values of the elemental Cu, Ag, and Au. 
All symmetrically independent single vacancies are considered.
}
\label{Cu-Ag-Au:Fig-Cu-Ag-Au-vacancy}
\end{figure}

Figure \ref{Cu-Ag-Au:Fig-Cu-Ag-Au-vacancy} (a) shows the excessive energy values required for forming a vacancy in the elemental Cu, Ag, and Au.
These values are obtained using polynomial MLP and the DFT calculation.
Figure \ref{Cu-Ag-Au:Fig-Cu-Ag-Au-vacancy} (b) also shows the formation energy values for the stable and metastable compounds containing single vacancies, which are measured from the energy values of the elemental Cu, Ag, and Au.
This figure shows only symmetrically independent single vacancies for each compound, with the same supercell size used in the phonon calculations.
The atomic positions and cell shape of the supercell with a single vacancy were fully optimized using the polynomial MLP and the DFT calculation.
The results indicate that the vacancy formation energy and the formation energy predicted using the polynomial MLP are similar to those obtained using the DFT calculation.

\section{Conclusion}
\label{Cu-Ag-Au:Sec-conclusion}

This study has demonstrated the development of polynomial MLP for the ternary Cu-Ag-Au alloy system.
The current MLP has been developed using a broad range of crystal structures, including local minimum structures obtained from global structure searches, covering the entire range of alloy compositions. 
The MLP shows strong capability in conducting global structure searches across various compositions in the ternary Cu-Ag-Au system and in predicting diverse properties for both binary and ternary compounds.

Developing a polynomial MLP with high accuracy in ternary alloy systems requires a large number of DFT calculations and complex models with numerous regression coefficients. 
Nevertheless, the developed MLP remains computationally efficient for evaluating energy, force, and stress tensor values of given structures. 
The methodology employed in this study can be readily adapted to develop MLPs for other alloy systems, facilitating robust global structure searches and atomistic simulations.

\begin{acknowledgments}
This work was supported by a Grant-in-Aid for Scientific Research (B) (Grant Number 22H01756) from the Japan Society for the Promotion of Science (JSPS).
\end{acknowledgments}

\bibliography{Cu-Ag-Au}

\end{document}